\documentstyle[12pt,epsfig]{article}
\def\bge{\begin{equation}}
\def\ene{\end{equation}}
\def\bg{\begin{eqnarray}}
\def\en{\end{eqnarray}}
\def\nn{\nonumber}

\def\ra{\rightarrow}

\begin{document}
\renewcommand{\thefootnote}{\fnsymbol{footnote}}
\begin{flushright}
ADP-00-27/T410\\
PCCF RI 00-12  
\end{flushright}
\vspace{0.5cm}
\begin{center}
{\LARGE Deep inelastic scattering on asymmetric nuclei}
\end{center}
\vspace{0.5cm}
\begin{center}
\begin{large}
K.~Saito$^1$\footnote{ksaito@nucl.phys.tohoku.ac.jp}, 
C. Boros$^2$\footnote{cboros@physics.adelaide.edu.au}, 
K. Tsushima$^2$\footnote{ktsushim@physics.adelaide.edu.au}, 
F. Bissey$^{2,4}$\footnote{fbissey@physics.adelaide.edu.au}, \\
I.R. Afnan$^3$\footnote{I.R.Afnan@flinders.edu.au} and 
A.W. Thomas$^2$\footnote{athomas@physics.adelaide.edu.au} \\
\end{large}
\vspace{0.5cm}
$^1$ Tohoku College of Pharmacy, Sendai 981-8558, Japan \\
$^2$ Department of Physics and Mathematical Physics, and \\
Special Recearch Center for the Subatomic Structure of Matter, \\
University of Adelaide, Adelaide 5005, Australia \\
$^3$  School of Chemistry Physics and Earth Sciences, Flinders University, \\
GPO Box 2100, Adelaide 5001, Australia \\
$^4$ Laboratoire de Physique Corpusculaire, Universit\'e Blaise Pascal, \\
CNRS/IN2P3, 24 avenue des Landais, 63177 Aubi\`ere Cedex, France 
\end{center}
\vspace{1cm}
\begin{abstract}
We study deep inelastic scattering on isospin asymmetric nuclei.  In 
particular, the difference of the nuclear structure functions and the 
Gottfried sum rule for the lightest mirror nuclei, $^3$He and $^3$H, are 
investigated. It is found 
that such systems can provide significant information on charge 
symmetry breaking and flavor asymmetry in the nuclear medium.  Furthermore, 
we propose a new method to extract the neutron structure function from 
radioactive isotopes far from the line of stability.  We also 
discuss the flavor asymmetry in the Drell-Yan process with isospin  
asymmetric nuclei.  
\end{abstract}
\vspace{0.5cm}

\newpage

The distributions of quarks in a nucleus differ significantly from 
those in the free nucleon.  This became clear when, nearly two decades ago, 
the European Muon Collaboration (EMC) measured the ratio of structure 
functions of iron and deuterium (D) in the deep inelastic scattering (DIS) 
of muons~\cite{emc}. Since then many experiments have been performed 
to investigate the distributions of partons in nuclei~\cite{review1}, and 
a large amount of data on (flavor) singlet nuclear structure functions 
has been collected.  
However, relatively less attention has been paid to the nonsinglet structure 
functions of a nucleus.  Measurements of the nonsinglet part of  nuclear 
structure functions could shed light on many phenomena involving 
non-perturbative QCD in {\em nuclei}, such as SU(3) symmetry breaking, 
the flavor asymmetry and so on.  

It is therefore very interesting and important to measure the structure 
functions of nuclei with large asymmetry in proton (p) and neutron (n) 
numbers ($Z \neq N$).  In particular, if DIS with high energy electrons  
on neutron or proton rich nuclei (for example, the 
RIKEN-MUSES project~\cite{muses}) 
could be realized in the future, it would be possible to get interesting 
information on the non-perturbative quark structure of nuclei.  In this paper, 
we study structure functions of nuclei with $Z \neq N$, including 
the radioactive isotopes (RI) which will be studied
at RIKEN~\cite{muses,tani}.  

Until recently it was usually assumed (see however Ref. \cite{Thomas:1983fh}) 
that the sea quark distributions 
in the (free) nucleon were flavor symmetric.  However, 
the New Muon Collaboration (NMC) at CERN~\cite{nmc} and the E866 Drell-Yan 
experiments at Fermilab~\cite{dyexp}, in particular,
have revealed a clear asymmetry in 
the sea distributions by using  proton and deuteron targets.  
If the quark and antiquark pairs in the sea are created perturbatively 
and charge symmetry breaking (CSB) effects ($m_d > m_u \sim$ 5MeV) 
are very small~\cite{csaba} in the free nucleon, it is not possible to 
explain the measured flavor asymmetry. 
This fact strongly suggests that  
non-perturbative QCD process, such as chiral symmetry breaking or the 
Pauli exclusion principle, must play an important role in the flavor
asymmetry~\cite{review2}.  

The flavor asymmetry in the sea was observed in 
the difference between the structure functions of the free proton and 
neutron, $F_2^p - F_2^n$, and it is simply expressed in terms of 
the $u$ and $d$ quark distributions, $u(x)$ and $d(x)$, and the 
anti-quark distributions, ${\bar u}(x)$ and ${\bar d}(x)$, in the free  
proton as   
\bge
F_2^p(x) - F_2^n(x) =
\frac{1}{3} x [u(x) - d(x)] + \frac{1}{3} x [{\bar u}(x) - {\bar d}(x)] ,   
\label{f2m}  
\ene
where we assume charge symmetry in the parton distributions~\cite{csaba}. 
(The structure functions and distributions depend on 
the 4-momentum transfer squared, $Q^2$, which is 
suppressed here.)  Then, the difference 
gives the Gottfried sum, $I_G^N$, for the nucleon as  
\bge
I_G^N(z) = \int_z^1 dx \frac{1}{x} [F_2^p(x) - F_2^n(x)] .  
\label{Ign1} 
\ene
Using the normalization condition for the valence quark distributions, 
we finally find 
\bge
I_G^N(0) = \frac{1}{3} + \frac{2}{3} \int_0^1 dx 
[{\bar u}(x) - {\bar d}(x)] ,  
\label{Ign2}
\ene
which provides a measure of flavor asymmetry.  
If the sea were flavor symmetric, namely 
${\bar u}(x) = {\bar d}(x)$, the Gottfried sum rule (GSR) is given 
by $I_G^N(0) = 1/3$.  However, the experimental value measured by 
NMC was $I_G^N(0) = 0.235 \pm 0.026$ (at $Q^2 = 4$ GeV$^2$)~\cite{nmc}, 
which is clearly less than $1/3$, and implies that ${\bar d}(x)$ 
overcomes ${\bar u}(x)$ in the free proton~\cite{nmc,dyexp}:  
\bge
\int_0^1 dx [{\bar d}(x) - {\bar u}(x)] = 0.148 \pm 0.039 . 
\label{Ignexp1}
\ene

The origin of the flavor asymmetry in the proton may be attributed 
to either the pion cloud required by 
chiral symmetry~\cite{Thomas:1983fh,piGott,tony2} or to the 
Pauli-exclusion principle at quark 
level~\cite{field,Signal:1989yc} (or both) -- 
since the proton consists of $uud$ quarks, $u{\bar u}$ pair creation 
may be suppressed by the Pauli-exclusion effect more 
than $d{\bar d}$ pair creation.  

It is a very interesting problem to see how the GSR changes in a nucleus.  
Studying the flavor asymmetry of a bound nucleon may provide an opportunity 
to learn details about the non-perturbative structure of the nucleon 
inside nuclear matter.   To this end we consider DIS of 
an electron or muon off {\em a pair of mirror nuclei}.   
We consider the ``nuclear Gottfried sum'' defined by 
\bge
I_G^{AA'}(z) = \int_z^A \frac{dx}{x} [F_2^A(x) - F_2^{A'}(x)] ,
\label{ngfsr1}
\ene
where ($A$, $A'$) is a pair of mirror nuclei: $A = Z + N (Z > N)$ 
(proton rich) and $A' = Z' + N' (N' > Z')$ (neutron rich).  

\begin{figure}[tb]
\begin{center}
\epsfig{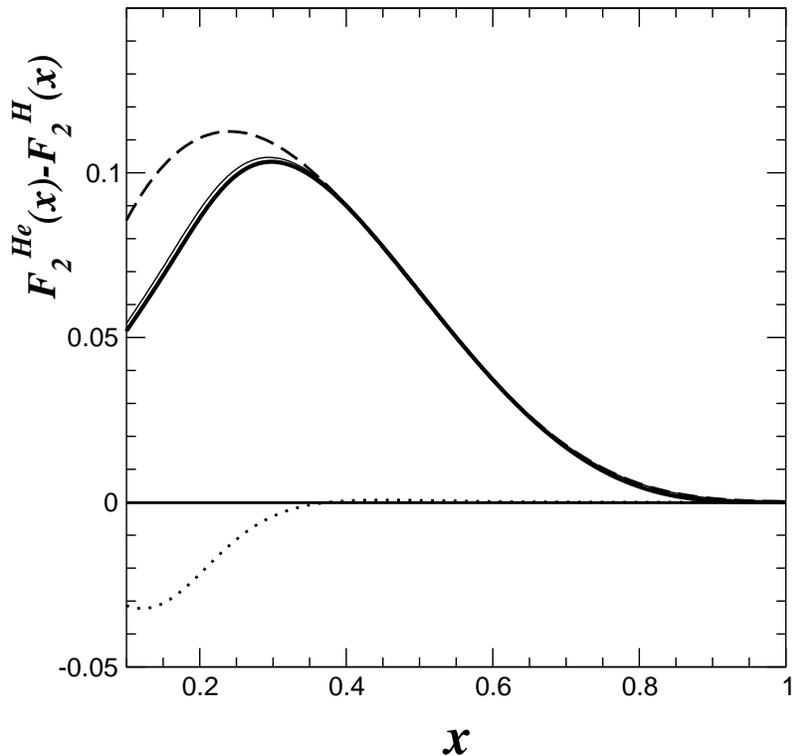}
\caption{Difference between the nuclear structure functions of $^3$He 
and $^3$H (at $Q^2 = 10$ GeV$^2$). The solid (dashed) [dotted] 
curves show the total (valence) [sea] contributions.   
The thin, solid curve presents the difference between the free 
proton and neutron structure functions (CTEQ5~\protect\cite{cteq5}). 
}
\label{fig:f2he}
\end{center}
\end{figure}
We now turn to the simplest example of this kind, namely the lightest 
mirror nuclei, $^3$He and $^3$H. 
The fact that these are mirror nuclei means that the nuclear corrections 
are very similar and this has already been exploited in a proposal to 
extract $F_{2p}/F_{2n}$ from a measurement of $F_2(^3{\rm 
He})/F_2(^3{\rm H})$ \cite{Afnan:2000uh}. 
The three-body system offers the advantage that it can 
be treated exactly by solving the Faddeev 
equations~\cite{review1,body3}. However, since it is not easy to handle the 
Coulomb interaction in the full Faddeev approach, we treat it approximately 
by adjusting the strength of the $^1S_0$ nucleon-nucleon potential so as to fit 
the observed binding energies: $E_b = - 7.72$ and $ - 8.48$ MeV for 
$^3$He and $^3$H, respectively~\cite{approx}.   

Using the spectral function given by the solution of the Faddeev equation 
with the PEST potential~\cite{pest},  
we have calculated the difference between the nuclear structure functions 
of $^3$He and $^3$H (at $Q^2 = 10$ GeV$^2$), which is shown 
in Fig.~\ref{fig:f2he}. For the free nucleon structure function 
we have used the CTEQ5 parametrization  
which incorporates the measured flavor asymmetry in
the free nucleon~\cite{cteq5}. We do not show  
the structure functions below $x = 0.1$ 
because in the small $x$ region we should include nuclear 
shadowing~\cite{shadow}.
Since we want to keep the present discussion simple in order to
emphasize the key physics ideas, for the present we concentrate on the nuclear 
structure functions in the region $x > 0.1$.
We may expect that since the difference between the
nuclear structure functions is (isovector) nonsinglet
the nuclear shadowing effect should not be 
large -- the pomeron does not directly couple to the isovector,
nonsinglet part. 

The effect of CSB in nuclei~\cite{miller} is to break the {\em nuclear} 
Gottfried sum rule (NGSR): 
\bge
I_G^{AA'}(0) =  Y  \times I_G^N(0) ,
\label{ngfsr2}
\ene
with $Y (= Z - N = N' - Z')$ the excess proton number in $A$.  
(Eq.(\ref{ngfsr2}) holds {\em if the nuclear environment does not 
affect the flavor asymmetry at all}.)  
%
%
%
%
In the present calculation we find that, at $Q^2 = 10$ GeV$^2$,
$I_G^N(0.1) = 0.152$ 
(from CTEQ5~\cite{cteq5}), while  $I_G^{He,H}(0.1) = 0.150$, which may imply
that CSB in the $A=3$ system gives very little change in the distribution of 
${\bar d}-{\bar u}$.  (See also Ref.~\cite{mol}.) 

While the major source of CSB is the Coulomb force, 
the main physics interest is in the possible additional 
sources of flavor asymmetry in nuclei, such as:
\begin{enumerate}
\item the probability of finding pions in the in-medium nucleon may be  
different from that in free space.  
It is certainly natural to expect that hadronic properties should change
in a nuclear medium~\cite{qmc,prop}. For example, 
the distribution of virtual pions per
bound nucleon may change, leading to a different flavor asymmetry in
nuclei~\cite{pion}.  The nuclear 
structure functions of mirror nuclei far from the line of stability  
may provide us significant information on flavor 
asymmetry and pion dynamics in nuclei. There will be many possible 
candidates for such mirror nuclei at facilities such as the proposed
radioactive ion beam collider at RIKEN~\cite{muses,tani}: 
for example, ($^{22}$Mg, $^{22}$Ne, $Y=2$), ($^{13}$O, $^{13}$B, $Y=3$), 
($^{17}$Ne, $^{17}$N, $Y=3$), ($^{20}$Mg, $^{20}$O, $Y=4$) etc.   
Those neutron rich or proton rich nuclei 
give rise to a large asymmetry in the $N/Z$ ratio.  
\item there is the much anticipated possibility of finding exotic
configurations, such as 6-quark states, in nuclei. As one consequence,
the effect of the Pauli exclusion principle at the quark level may be 
different because of quark percolation. 
In the naive counting estimate the ground state of the free proton has 
4 vacant states for $u$ quarks and 5 vacant states for $d$ quarks.  
Thus, the expected flavor asymmetry is 
${\bar d}/{\bar u} \simeq 5/4$~\cite{field,Signal:1989yc}.  
However, in a nuclear medium there may be some probability for two 
nucleons to overlap.  Suppose that the valence quarks from two protons 
are put into one confining potential. In this case,   
there are 2 vacant states (in the lowest energy level) 
for $u$ quarks and 4 vacant states for $d$ quarks.   
Then, the asymmetry becomes ${\bar d}/{\bar u} \simeq 2$, which implies 
that the flavor asymmetry would be enhanced in nuclei.  
It may be possible to investigate these nonperturbative phenomena by 
measuring the nuclear structure functions of mirror nuclei.  
\end{enumerate}

DIS off mirror nuclei with large isospin asymmetry should be possible  
in the future. Our calculation for $A=3$ represents the standard
calculation without any of the possible new contributions. 
The observation of some deviation 
from the calculations shown here would stimulate a great deal of work
which may eventually lead to genuinely new information on the dynamics
of nuclear systems.  

We would like to add some comments on isospin 
asymmetric nuclei (or nuclei far from the line of stability).   First, 
it is very important to extract the neutron structure function from  
nuclear experimental data. The system traditionally used for this is
the deuteron but the A=3 case is also being seriously considered now 
\cite{Afnan:2000uh}. 
The proposed MUSES project at RIKEN~\cite{muses} will provide  
a number of attractive alternatives for extracting
the neutron structure function using radioactive ions rich in neutrons.  

As an example, let us consider a series of Li nuclei.  There exist five  
RI observed at RIKEN: $^6$Li, $^7$Li, $^8$Li, $^9$Li, $^{11}$Li.  
Then, we shall consider a ratio  
\bge
R_n(i) = \frac{F_2(^{i}\mbox{Li}) - F_2(^{i-1}\mbox{Li})}{F_2^D - F_2^p} ,  
\label{nrich} 
\ene 
where $F_2(^{i}\mbox{Li})$ is the structure function of $^i$Li, and 
$i$ runs from 7 to 9.  (For $i = 11$, we may take  
$[F_2(^{11}\mbox{Li}) - F_2(^{9}\mbox{Li})]/2$ as the numerator of  
the ratio.)  This $R_n(i)$ means the ratio of the neutron structure  
function bound in the Li nucleus to that in D (if those were  
the same the ratio would be unity).  We suppose that  
the deuteron structure function itself is already understood.  

Varying the atomic number $i$ successively (i.e., varying the binding  
effect on the neutron in Li) and comparing the $i$-series of those ratios, 
we may get useful information on the neutron structure function, 
its nuclear binding effect (shell structure dependence)~\cite{tani}  
and the off-mass shell effect~\cite{shadow}.  
In particular, the last neutron in the halo nuclei (like $^{11}$Li) is  
bound very weakly -- some hundred keV~\cite{muses,tani}, which is much  
weaker than in D.  The RI series of Be, B, C, O, Na and 
Mg~\cite{muses,tani} are also of great interest. 

{}Finally we turn to the nuclear Drell-Yan (DY) process as 
an alternative method to 
study the sea quark distributions~\cite{kumano}.  
We first define the $p$-$n$ asymmetry, $C_{DY}^N$, in the free nucleon as    
\bge
C_{DY}^N = \frac{2\sigma^{pp}-\sigma^{pD}}{\sigma^{pD}} 
 \simeq \frac{\sigma^{pp}-\sigma^{pn}}{\sigma^{pp}+\sigma^{pn}} , 
\label{cn}
\ene
where $\sigma^{AB}$ is the DY cross section for the lepton pair production 
of $A + B \ra \ell^+ \ell^- + X$, and we assume that the nuclear 
binding effect in D can be neglected. 
Then, if we take a large $x_F (= |x_A - x_B|)$ the $p$-$n$ asymmetry becomes 
\bge
C_{DY}^N \ra \frac{[4u(x_A) - d(x_A)][{\bar u}(x_B) - {\bar d}(x_B)]}{
[4u(x_A) + d(x_A)][{\bar u}(x_B) + {\bar d}(x_B)]} , 
\label{cn1}
\ene
where $u(x_i)$ ($d(x_i)$) is the $u$ ($d$) quark distribution with momentum 
fraction $x_i$ in $i$ ($= A$ or $B$).  
This ratio is sensitive to the flavor asymmetry in $B$.  

In order to check 
the flavor asymmetry in a nucleus, we shall again use a pair of 
mirror nuclei, $(A, A')$, and define the $p$-$n$ asymmetry in nuclei as 
\bge
C_{DY}^{AA'} = \frac{\sigma^{pA}-\sigma^{pA'}}{\sigma^{pD}} .  
\label{cAA}
\ene
If $A$ and $A'$ simply consist of protons and neutrons and we assume that 
the nuclear binding effect is small, the $p$-$n$ asymmetry in nuclei becomes
\bge
C_{DY}^{AA'} = Y \times C_{DY}^N ,
\label{cAA1}
\ene  
with the excess proton number $Y$.  If we take the (double) ratio as 
\bge
D_{DY} = \frac{C_{DY}^{AA'}}{C_{DY}^N} = Y , 
\label{cAA2}
\ene
it gives just $Y$ if the flavor asymmetry is not changed by 
the nuclear medium.  The deviation from $Y$ in actual measurements should be 
a sensitive signature for the flavor asymmetry and nuclear binding 
in a nucleus.  

Moreover, we can see the flavor asymmetry by using a single 
RI.  We suppose a ratio 
\bge
R_{DY}^{A} = \frac{2}{A} \left( \frac{\sigma^{pA}}{\sigma^{pD}} 
\right) ,   
\label{rA}
\ene
and again assume that the nucleus, $A$, is a simple collection of $Z$ protons 
and $N$ neutrons.  We then find 
\bg
R_{DY}^A  
&\simeq& \frac{2}{A} \left( 
\frac{Z \sigma^{pp} + N \sigma^{pn}}{\sigma^{pp}+\sigma^{pn}} 
\right) , \nn \\
&\simeq& 1 - \frac{(N-Z)}{A} \left( 
\frac{\sigma^{pp} - \sigma^{pn}}{\sigma^{pp}+\sigma^{pn}} \right) , \nn \\
&\sim& 1 - \frac{(N-Z)}{A} \cdot 
\frac{[4u(x_A) - d(x_A)][{\bar u}(x_B) - {\bar d}(x_B)]}{
[4u(x_A) + d(x_A)][{\bar u}(x_B) + {\bar d}(x_B)]}  ,  
\label{rA1}
\en
for large $x_F$.  
This is also sensitive to the flavor asymmetry in the nucleus.  
In particular, nuclei far from the line of stability 
are very interesting because 
they give large asymmetries -- i.e., large values of $(N-Z)/A$.  

In summary, we have studied the nuclear structure functions and the 
Gottfried sum rule for a pair of mirror nuclei which could give very 
significant information on CSB and flavor asymmetry in the nuclear 
medium.  For the present we have restricted our study to the region 
$x > 0.1$,  
in order to make the discussion simple.  It will be very interesting to 
extend this work to calculate the nuclear structure functions including 
shadowing, in order to 
complete the discussion of the NGSR and flavor 
asymmetry in nuclei.  Furthermore, we have proposed a new method 
to extract the neutron structure function from RI far from 
the line of stability, 
and discussed the flavor asymmetry in the nuclear 
DY process.  If DIS with high energy electrons 
(or muons) off nuclei with large isospin asymmetry 
becomes possible in the future, 
we could get a lot of valuable information on the non-perturbative quark 
structure of atomic nuclei.  

\vspace{0.5cm}

We would like to thank W. Melnitchouk for many useful 
discussions.  K. S. would also like to thank J.T. Londergan for providing   
a lot of useful information on flavor asymmetry.   
This work was partly supported by the Australian Research 
Council.  

\newpage


\begin{thebibliography}{99}
%
\bibitem{emc} J.J. Aubert et al. (EMC collaboration), Phys. Lett. B 123 
(1983) 275.  
%
\bibitem{review1} D.F. Geesaman, K. Saito, A.W. Thomas, Annu. Rev. Nucl. 
Part. Sci. 45 (1995) 337; M. Arneodo, Phys. Rep. 240 (1994) 301.  
%
\bibitem{muses} MUSES (Multi-USe Experimental Storage-rings) project at  
RIKEN, see the RIKEN home page (http://www.rarf.riken.go.jp).  
%
\bibitem{tani} I. Tanihata, Nucl. Phys. A 654 (1999) 235c.  
%
\bibitem{Thomas:1983fh} A.W. Thomas, Phys. Lett. B 126 (1983) 97. 
%
\bibitem{nmc} M. Arneodo et al. (New Muon collaboration), Phys. Rev. D 50 
(1994) R1. 
%
\bibitem{dyexp} J.C. Peng et al. (FNAL E866/NuSea collaboration), 
Phys. Rev. D 58 (1998) 092004.  
%
\bibitem{csaba} C. Boros, F.M. Steffens, J.T. Londergan, A.W. Thomas,
Phys. Lett. B 468 (1999) 161; J.T. Londergan, A.W. Thomas, 
Prog. Part. Nucl. Phys. 41 (1998) 49.
%
\bibitem{review2} S. Kumano, Phys. Rep. 303 (1998) 183; 
J. Speth, A.W. Thomas, Adv. in Nucl. Phys. 24 (1998) 83; 
J.C. Peng, G.T. Garvey, hep-ph/9912370 (LA-UR-99-5003), to be published in 
{\it ``Trends in Particle and Nuclear Physics''}, vol.1 (Plenum Press, N.Y.). 
%
\bibitem{piGott}
E.M. Henley, G.A. Miller, Phys. Lett. B 251 (1990) 453; 
S. Kumano, J.T. Londergan, Phys. Rev. D 44 (1991) 717; 
A. Signal, A.W. Schreiber, A. W. Thomas, Mod. Phys. Lett. A 6 (1991) 271; 
M. Ericson, A.W. Thomas, Phys. Lett. B 148 (1984) 191. 
%
\bibitem{tony2} A.W. Thomas, W. Melnitchouk, F.M. Steffens, hep-ph/0005043.
%
\bibitem{field} R.D. Field, R.P. Feynman, Phys. Rev. D 15 (1977) 2590. 
%
\bibitem{Signal:1989yc} A.I. Signal, A.W. Thomas, Phys. Rev. D 40 (1989) 
2832. 
%
\bibitem{Afnan:2000uh}
I.R. Afnan, F. Bissey, J. Gomez, A.T. Katramatou, W. Melnitchouk, 
G.G. Petratos, A.W. Thomas, nucl-th/0006003.
%
\bibitem{body3} T. Uchiyama, K. Saito, Phys. Rev. C 38 (1988) 2245; 
C. Ciofi degli Atti, S. Liuti, Phys. Rev. C 41 (1990) 1100. 
%
\bibitem{approx} F. Bissey, A.W. Thomas, I.R. Afnan, manuscript in 
preparation. 
%
\bibitem{pest}  J. Haidenbauer, W. Plessas, Phys. Rev. C 30 (1984) 1822.
%
\bibitem{cteq5} H.L. Lai et al., Eur. Phys. J. C 12 (2000) 375.
%
\bibitem{shadow} For example, G. Piller, W. Weise, Phys. Rep. 330 
(2000) 1. 
%
\bibitem{miller} For example, G.A. Miller, W.T.H. Van Oers, 
{\it ``Symmetries and Fundamental Interactions in Nuclei''}, edited by
W.C. Haxton and E.M. Henley, World Scientific (1995), p. 127. 
%
\bibitem{mol} V.V. Burov, A.V. Molochkov, G.I. Smirnov, Phys. Lett. 
B 466 (1999) 1. 
%
\bibitem{qmc} P.A.M. Guichon, K. Saito, E. Rodionov, A.W. Thomas,  
Nucl. Phys. A 601 (1996) 349; K. Saito, K. Tsushima, A.W. Thomas, 
Nucl. Phys. A 609 (1996) 339;  K. Saito, K. Tsushima, A.W. Thomas, 
Phys. Rev. C 55 (1997) 2637. 
%
\bibitem{prop} Quark Matter '99, Nucl. Phys. A 661 (1999).
%
\bibitem{pion} M. Ericson, A.W. Thomas, Phys. Lett. B 128 (1983) 112; 
E.L. Berger, F. Coester, Phys.  Rev. D 32 (1985) 1071; 
C.L. Korpa, A.E.L. Dieperink, Phys. Lett. B 446 (1999) 15. 
%
\bibitem{kumano} See also, S. Kumano, Phys. Lett. B 342 (1995) 339.  
%
\end{thebibliography}
\end{document}